\documentstyle[12pt]{article}

\renewcommand{\theequation}{\thesection.\arabic{equation}}

\newcommand \beq{\begin{eqnarray}}
\newcommand \eeq{\end{eqnarray}}
\def\bfepsilon{\mbox{\boldmath$\epsilon$}}

\def\del{\partial}                              

\def\frac#1#2{{#1 \over #2}}

\def\half{\ifinner {\scriptstyle {1 \over 2}}
   \else {1 \over 2} \fi}

\def\simge{\mathrel{%
   \rlap{\raise 0.511ex \hbox{$>$}}{\lower 0.511ex \hbox{$\sim$}}}}
\def\simle{\mathrel{
   \rlap{\raise 0.511ex \hbox{$<$}}{\lower 0.511ex \hbox{$\sim$}}}}

\def\slashchar#1{\setbox0=\hbox{$#1$}           
   \dimen0=\wd0                                 
   \setbox1=\hbox{/} \dimen1=\wd1               
   \ifdim\dimen0>\dimen1                        
      \rlap{\hbox to \dimen0{\hfil/\hfil}}      
      #1                                        
   \else                                        
      \rlap{\hbox to \dimen1{\hfil$#1$\hfil}}   
      /                                         
   \fi}                                         %

\def\subrightarrow#1{
  \setbox0=\hbox{
    $\displaystyle\mathop{}
    \limits_{#1}$}
  \dimen0=\wd0
  \advance \dimen0 by .5em
  \mathrel{
    \mathop{\hbox to \dimen0{\rightarrowfill}}
       \limits_{#1}}}                           

\def\journal#1#2#3#4{\ {#1}{\bf #2} ({#3})\  {#4}}

\def\AnnPhys{\journal{Ann.\ Phys.}}

\def\NPB{\journal{Nucl.\ Phys.\ {\bf B}}}

\def\PRD{\journal{Phys.\ Rev.\ {\bf D}}}
\def\PRL{\journal{Phys.\ Rev.\ Lett.}}

\def\SovPhysJETP{\journal{Sov.\ Phys.\ JETP}}

\def\SovJNuclPhys{\journal{Sov.\ J.\ Nucl.\ Phys.}}

\begin{document}

\begin{titlepage}
\begin{flushright}

{Saclay-T94/03}
\end{flushright}
\vspace*{3cm}
\begin{center}
\baselineskip=13pt
{\Large ENERGY-MOMENTUM TENSORS\\ FOR THE QUARK-GLUON PLASMA\\}
\vskip0.5cm
Jean-Paul BLAIZOT\footnote{CNRS}  and
Edmond IANCU\\
{\it Service de Physique Th\'eorique\footnote{Laboratoire de la Direction des
Sciences de la Mati\`ere du Commissariat \`a l'Energie
Atomique}, CE-Saclay \\ 91191 Gif-sur-Yvette, France}\\
\vskip0.5cm
January 1994

\vskip 2cm \begin{abstract}
We construct the energy-momentum tensor for the  gauge fields
which describe the collective  excitations of the quark-gluon plasma.
We rely on the description of the collective modes that we have derived in
 previous works. By using the conservation laws for  energy and momentum,
we obtain three different versions for the tensor $T^{\mu\nu}$, which
are physically equivalent. We show that the total energy constructed
from $T^{00}$ is positive for any non-trivial field configuration.
Finally, we present a new non-abelian solution of the equations
of motion for the gauge fields. This solution corresponds to spatially uniform
 color oscillations of the plasma.
\end{abstract}
\vskip 1cm
\begin{flushleft}
Submitted to Nuclear Physics B
\end{flushleft}
\end{center}

\end{titlepage}

\setcounter{equation}{0}
\section{Introduction}

The low-lying excitations of the quark-gluon plasma at high temperature
are collective  excitations which involve the coherent behaviour of a  large
number of particles over typical distances and times of order $1/gT$.
These  excitations  can
 be conveniently described as oscillations of  self-consistent average
 fields to which the plasma particles couple\cite{us}.
We have shown that, at leading order in the coupling $g$, the
 Dyson-Schwinger
 equations for the $N$-point functions ($N\ge 1$) reduce to a set of
coupled equations for these mean fields and for 2-point functions.
 The equations for the 2-point functions
describe the dynamics of the hard plasma particles in the presence of soft
background fields; they may be given the form of
 simple kinetic equations for generalized on-shell distribution
functions\cite{us}.

In the present work, we  investigate further the  properties of the
collective modes, in order to properly characterize their energy-momentum
content. We restrict ourselves here to the case of {\it bosonic}
collective excitations.
Special attention has been payed recently to this
problem\cite{Weldon92,Brandt93,Nair93}. An early attempt to derive
the corresponding energy-momentum tensor $T^{\mu\nu}$ can be found in
Ref.~\cite{Weldon92}. There, auxiliary fields are introduced in order
to write a local effective action, for which the Noether construction
is available. The expression for  $T^{\mu\nu}$ thus obtained is rather
intricate and difficult to use.
Another approach is used in Ref.~\cite{Brandt93},
where the plasma is coupled with a weak gravitational field.
Again, a complicated form is obtained for $T^{\mu\nu}$, which makes its
physical interpretation difficult.
 This is true, in particular, for the terms quadratic in
  the gauge fields  $A_{\mu}$, which, however, should be analogous to those
corresponding to an abelian plasma.
We would certainly expect the expressions appropriate for
 the QED plasma to be simple, easy to interpret, and
closely related to the ones corresponding to classical polarizable media.
Another difficulty with the  derivation proposed in \cite{Brandt93}
is that  the resulting field energy, $P^0\equiv \int d^3x T^{00}$,
appears to be negative
even for some simple, plane-wave, field configurations, thus casting doubts
on the stability of the equilibrium state.
It is, however, verified in \cite{Brandt93}
 that $P^0$  is positive for the  abelian-like plasma normal modes.
A similar, but more general, conclusion emerges from Ref.~\cite{Nair93}, where
a Hamiltonian for the
soft gauge fields is constructed in a completely
different approach\cite{Efraty92}, by exploiting the analogy between
the effective action for the collective excitations\cite{Wong90,Braaten92,us}
 and the eikonal of a Chern-Simons theory. The resulting functional is positive
when evaluated for gauge fields which satisfy the non-abelian equations of
motion without sources, that is, for general, non-abelian, collective
excitations.

As we shall show, our approach overcomes most of the difficulties mentioned
above and has the advantage of offering a transparent physical interpretation.
We study a general configuration of gauge fields which
 may be induced by an appropriate external color source
$j^\mu_a$. By using the
conservation laws for energy and momentum, we construct the tensor
 $T^{\mu\nu}$ for such gauge fields.
 In this procedure, we rely on the explicit expressions
 for the retarded non-abelian induced current $j_\mu^{ind}$
that we have obtained in previous works\cite{us}. The essential aspects
of our approach are presented in Section 2.

Of course, the construction of  $T^{\mu\nu}$  from conservation laws
does not give a unique answer, and, in fact, we shall
 write down three different expressions,
which are physically equivalent:\\
({\bf i}) The first expression, derived in Sec.3.1. and referred simply
as  $T^{\mu\nu}$ generalizes
 well-known expressions for  classical dielectric media
(see, e.g., Refs.\cite{Landau}), to which it reduces in the abelian case.
The corresponding energy density $T^{00}$ is {\it manifestly positive}
for arbitrary gauge fields. The same is therefore true for the corresponding
field energy $P^0$.\\
({\bf ii}) The second version for the energy-momentum tensor, referred as
$\tilde T^{\mu\nu}$, is derived in Sec.3.2. It is equivalent,
but slightly simpler, to the  expression obtained by Weldon
 in Ref.~\cite{Weldon92},
and differs from it only through a ``superpotential''\cite{Callen}.
The components $\tilde T^{\mu i}$ ($i=1,\,2,\,3$)
of this new tensor have a  simpler non-local structure
than  $T^{\mu i}$.\\
({\bf iii}) Finally, the third version, derived in Sec.3.3 and denoted
as $T^{\mu\nu}_{{\rm sym}}$, is traceless and  symmetric. This
 allows us, in particular, to construct the gauge field angular momentum.

In Section 4, we show that the expressions that we have obtained reduce to
familiar ones when specialized to the case of abelian or quasi-abelian
plasmas (i.e., in the weak field limit). In Section 5, we briefly discuss
new, truly non-abelian, solutions of the equations of motion, which
correspond to global color oscillations. The main conclusions are summarized
 in Section 6.
\setcounter{equation}{0}
\section{Soft fields and hard particles}

  We consider an ultrarelativistic quark-gluon plasma
 close to thermal equilibrum, at a temperature $T=1/\beta$ and zero
chemical potential.
We use  natural units, $\hbar=c=1$, and  Minkovski metric.
We consider a $SU(N)$ gauge theory with $N_f$ flavors of quarks.
The generators of the gauge group in  different representations
are taken to be Hermitian and traceless. They are denoted by $t^a$ and
$T^a$, respectively, for the fundamental and the adjoint representation,
and are normalized such that $tr(t^a t^b)=(1/2)\delta^{ab}$ and
 $tr(T^aT^b)=N\delta^{ab}$.
It follows that ($T^a)_{bc}=-if^{abc}$, where $f^{abc}$ are the structure
constants of the group: $[T^a,T^b]=if^{abc}T^c.$
We use, without distinction, upper and lower positions for the color
indices.

As alluded to in the introductory section, we are studying
the response of the plasma to a  classical, external,
 color current $j_\mu\equiv j_\mu^a \,T^a$.
In the absence of the external source, the plasma is assumed to be in
thermal equilibrium and the expectation values of the fields  vanish.
The interaction with  the external  current is adiabatically turned on
from time $t_0\to -\infty$, and the gauge fields
 acquire then nontrivial  expectation values, denoted by $A_\mu^a(x)$.

We assume the external perturbation  to be {\it weak} and {\it slowly varying}.
Then, the induced average fields have Fourier components with
 only {\it soft} momenta $P\sim gT$,
and  their amplitude is constrained so that $F_{\mu\nu}
\simle gT^2$, or, equivalently, $A_\mu\simle T$\cite{us}.
The latter limitation on the amplitude of the field oscillations
ensures in particular  the consistency of the soft  covariant derivatives:
if $A_\mu\sim T$ , then $g A_\mu\sim gT$
is of the same order as the derivative of a slowly varying
quantity, $i\del_\mu\sim g A_\mu$.
Occasionally, we shall  specialize our results to the case of weaker
fields, for which
 it is consistent to replace covariant derivatives by ordinary
ones; then, $A_\mu$ behaves as an abelian gauge field and the equations
are linear. This weak field limit  will be referred
below as ``the abelian regime''. (For a
QED plasma, the equations remain linear even for fields as strong
as allowed, i.e., for $A_\mu\sim T$.)

The gauge fields  satisfy the generalized Maxwell equation
\beq
\label{ava}
\left [\, D^\nu,\, F_{\nu\mu}(x)\,\right ]^a
\,=\,j_\mu^a(x)+j_\mu^{ind\, a}(x),
\eeq
where $D_\mu = \del_\mu+igA_\mu(x),$ ($A_\mu\equiv A_\mu^a T^a$),
and  $F_{\mu\nu}= [D_\mu, D_\nu]/(ig) = F_{\mu\nu}^aT^a$.
The external source $j^a_\mu(x)$ is to be considered as a formal
device to generate arbitrary gauge field configurations.
In leading order, the induced current
 $j_\mu^{ind}$  is proportional to the fluctuations
of the average  phase-space color  densities of  plasma constituents. These
 fluctuations are described by color matrices which
we denote by $\delta n_\pm({\bf k},x)\equiv\delta n_\pm^a({\bf k},x)\,t^a$
 and $\delta N({\bf k},x)\equiv \delta N^a({\bf k},x)\,T^a$ for
quarks, antiquarks and  gluons, respectively. Then
\beq\label{jind}
j_\mu^{ind\,a}(x)&=&g\int\frac{d^3k}{(2\pi)^3}\,v_\mu
\,{\rm Tr}\,\biggl\{2\,N_{\rm f}\,t^a\left[ \delta
 n_+({\bf k},x)-\delta n_-({\bf k},x)\right]\,+\,
2\,T^a\,\delta N({\bf k},x)\biggr\}\nonumber\\
&=&g\int\frac{d^3k}{(2\pi)^3}\,\frac{k_\mu}{k}
\left\{\,N_{\rm f}\left[ \delta
 n_+^a({\bf k},x)-\delta n_-^a({\bf k},x)\right]\,+\,
2N\,\delta N^a({\bf k},x)\right\},\eeq
where $k^0=|{\bf k}|$ and the factors of 2 account for the spin degrees of
 freedom. The quantities  $\delta n_\pm$ and $\delta N$
are determined by the following kinetic
equations\cite{us}:
\beq\label{n}
\left[ v\cdot D_x,\,\delta n_\pm({{\bf k}},x)\right]=\mp\, g\,{\bf v}
\cdot{\bf E}(x)\,\frac{dn_k}{dk},\eeq
\beq\label{N}
\left[ v\cdot D_x,\,\delta N({{\bf k}},x)\right]=-\, g\,
{\bf v}\cdot{\bf E}(x)\frac{dN_k}{dk}.\eeq
Here, $v^\mu\equiv (1,\,{\bf v})$, where ${\bf v}\equiv
{\bf k}/k$ is the velocity of the hard particle ($k\equiv |{\bf k}|$),
 $E^i\equiv F^{i0}$ is the chromoelectric field.
In the r.h.s., $N_k
\equiv 1/(exp(\beta k)-1)$ and $n_k\equiv 1/(exp(\beta k)+1)$ denote,
 respectively, equilibrium boson and fermion occupation factors.

 Eqs.~(\ref{n}) and (\ref{N})
generalize the Vlasov equation to non-abelian plasmas.
They are covariant under a local gauge
transformation of  the mean fields $A^a_\mu(x)$; therefore, the fluctuations
 $\delta n_\pm^a({\bf k},x)$ and $\delta N^a({\bf k},x)$
 transform like vectors in the adjoint representation.
The gauge symmetry requires the presence of covariant derivatives, which
makes these equations non-linear with respect to $A^a_\mu$.
If we were to solve these equations  for a fixed
$v^\mu\equiv (1,\,{\bf v})$, we could get rid of the non-linear terms
by choosing the {\it axial gauge}  $v^\mu A^a_\mu(x)=0$.
In this gauge,
$(v\cdot D)^{ab}=\delta ^{ab}\,v\cdot\del$ and ${\bf v}\cdot{\bf E}^a(x)=
-\,{\bf v}\cdot (\del_0 {\bf A}^a+\nabla A_0^a)$, as for abelian fields.
However, in  calculating the  induced current (\ref{jind}) we have to
integrate over all the directions of ${\bf v}$.
It is therefore necessary to solve eqs.~(\ref{n})--(\ref{N}) in
an arbitrary gauge.  This can be done with the help of the
 retarded Green's functions for the covariant line derivative
$v\cdot D_x$, defined by
\beq\label{Gret}
i\,(v\cdot D_x)_{ac}\,G^{\,cb}_{ret}(x,y;v)=\delta^{ab}\,\delta^{(4)}(x-y),
\qquad G_{ret}(x,y;v)=0\,\,\,{\rm for}\,\,x_0<y_0,
\eeq
and  which has the following expression
\beq\label{GR}
G_{ret}(x,y;v)&=&-i\,\theta (x^0-y^0)\,\delta^{(3)}
\left({{\bf x}}-{{\bf y}}-{{\bf v}}(x^0-y^0)
\right )U(x,y).\eeq
Here,  $U(x,y)$  is the parallel transporter along the straight line
$\gamma$   joining $x$ and $y$,
\beq\label{pt} U(x,y)=P\exp\{ -ig\int_\gamma dz^\mu A_\mu(z)\}.\eeq
The solution of eq.~(\ref{n}) is then
\beq\label{nsol}
\delta n_\pm^a({\bf k},x)
&=&\mp\, g\,\frac {d n_k}{d k}\int_0^\infty du\, U_{ab}(x,x-vu)\,
{\bf v}\cdot{\bf E}^b(x-vu), \eeq
with a similar expression for $\delta N^a({\bf k},x)$.
{}From eqs.~(\ref{jind}) and (\ref{nsol}), we readily derive the induced
current
\beq\label{solj}
j^{ind}_{\mu\,a}(x)\,=\,3\,\omega^2_p\int\frac{d\Omega}{4\pi}
\,v_\mu\int_0^\infty du\, U_{ab}(x,x-vu)\, {\bf v}\cdot{\bf E}^b(x-vu),\eeq
where $\omega_p$ is the {\it plasma frequency},
$\omega^2_p\equiv (2N+N_{\rm f})g^2 T^2/18$. The  integral $\int
d\Omega$ runs over all the directions of the unit vector ${\bf v}$.
It can be easily verified that $j^{ind}_{\mu\,a}(x)$ is covariantly conserved,
\beq \label{jcons}\left[D^\mu,\,j_\mu^{ind}(x)\right]=0.\eeq
It follows that the external current is constrained by a similar equation,
$[D^\mu,\,j_\mu(x)]=0$.

Eq.~(\ref{solj}) for the induced current summarizes all the
information about the  collective motion of the hard particles in the presence
of the soft gauge fields. By using it in the r.h.s. of eq.~(\ref{ava}),  one
 obtains a generalization of  Maxwell's equations in a polarizable
 medium.  Besides , the induced current
 acts as a {\it generating functional for all the (retarded) amplitudes between
soft gauge fields}. For instance,  the  polarization tensor for soft gluons is
\beq\label{PI}
\Pi_{\mu\nu}^{ab}(x,y)=\frac{\delta j_{\mu\,a}^{ind}(x)}{\delta A_b^\nu(y)}
\Big |_{A=0},\eeq
which gives (with $P^\mu = (p^0,\,{\bf p})$)
\beq\label{P}
\Pi^{ab}_{\mu\nu}(P)= 3\,\omega_p^2\,\delta^{ab}
\left \{-\delta^0_\mu\delta^0_\nu \,+\,p^0 \int\frac{d\Omega}{4\pi}
\frac{v_\mu\, v_\nu} {v\cdot P+i\eta}\right\}.
\eeq This is the well-known ``hard thermal loop'' for the gluon
self-energy\cite{Silin60,Klimov81,Weldon82,Braaten90b}.
In the weak field limit, the tensor $\Pi_{\mu\nu}$  fully
characterizes the dielectric properties of the plasma
For strong fields, $A_\mu\sim T$, all $N$-gluons
amplitudes with $N\ge 2$ contribute to the induced current (\ref{solj}) at
the same order in $g$\cite{Pisarski,Braaten90b,us}.

The induced current may be given a different functional
 form which will be used later.
We start by defining a new function, $W^\mu(x;v)$, as the solution to
\beq\label{eqw}
\left[ v\cdot D_x,\, W^\mu(x;v)\right]\,=\,F^{\mu\rho}(x)\,v_\rho,\eeq
which vanishes when $x_0\to -\infty$. We have $W^\mu=W_a^\mu T^a$, with
\beq\label{W}
W^\mu_a(x;v)\,=\, \int_0^\infty du\, U_{ab}(x,x-vu)\, F_b^{\mu\rho}(x-vu)
\,v_\rho,\eeq
and the induced current (\ref{solj}) can be rewritten as
\beq\label{j1}
j^{ind}_{\mu\,a}(x)\,=\,3\,\omega^2_p\int\frac{d\Omega}{4\pi}
\,v_\mu\,W_a^0(x;v).\eeq
In the case of an abelian plasma,
$e W^\mu$ has a simple physical interpretation: it represents
the 4-momentum transferred by the electromagnetic field to a  particle
of charge $e$ and velocity ${\bf v}$, moving  along
 a straight line from time $t_0\to -\infty$ to $x_0$.
(The transferred momentum $\sim gT$ is small compared to the  hard
momentum $k\sim T$, so that the particle is not significantly deviated by
the Lorentz force.)
{}From eq.~(\ref{W}) it is obvious that
\beq\label{trans} v_\mu\,W^\mu_a(x;v)\,=\,0.\eeq
The functions $W^\mu_a(x;v)$ are homogeneous of degree zero as functions
of $v^\mu$, i.e.,
\beq\label{homo}
v^\nu\,\frac{\del W^\mu_a(x;v)}{\del v^\nu}\,=\,0.\eeq
(In taking derivatives with respect to $v^\mu$, as in eq.~(\ref{homo}),
we consider the four components of $v^\mu$ as independent variables, and
set $v^\mu=(1, {\bf v})$ only after derivation.)
This is obvious from the defining equation (\ref{eqw}), and can be also
verified on eq.~(\ref{W}) by using the identity
\beq v^\nu\,\frac{\del}{\del v^\nu}\,f(x-vu)\,=\,-u\,(v\cdot\del_x)f(x-vu).\eeq

We  prove now that the  expressions (\ref{solj}) or (\ref{j1})
for the induced current are equivalent to the following one
\beq\label{j2}
j^{ind}_{\mu\,a}(x)\,=\,\frac{3}{2}\,\omega^2_p\int\frac{d\Omega}{4\pi}
\,\left [ W_\mu^a(x;v)\,+\, v_\mu\,\frac{\del W_a^\rho(x;v)}
{\del v^\rho}\right ],\eeq
which also appears, with sligthly different notations, in the last of
Refs.\cite{us} (as eq.~(C.15)).
To show the equivalence, we
note first that eq.~(\ref{homo}) implies
\beq\label{del}
\frac{\del W_a^\rho(x;v)}{\del v^\rho}
\,=\,\frac{\del W_a^i(x;v)}{\del v^i}\,-\,\left (v^i \frac{\del}{\del v^i}
\right )W_a^0(x;v).\eeq
Then, we use eq.~(\ref{trans}) to replace $W_a^0(x;v)$
by ${\bf v}\cdot {\bf W}_a(x;v)$ in (\ref{del}), and obtain
\beq\label{totder}
\frac{\del W_a^\rho(x;v)}{\del v^\rho}\,=\,W_a^0(x;v)\,+\,
\left ( \frac{\del}{\del v^i}-\frac{\del}{\del v^j}\,\frac
{v^i v^j}{{\bf v}^2} \right )W_a^i(x;v).\eeq
The second term in the r.h.s. vanishes upon angular integration,
so that
\beq
\int\frac{d\Omega}{4\pi}\,\frac{\del W_a^\rho(x;v)}{\del v^\rho}\,=\,
\int\frac{d\Omega}{4\pi}\,W_a^0(x;v).\eeq
This proves the equivalence between the time components ($\mu =0$) of
eqs.~(\ref{j1}) and (\ref{j2}). The corresponding proof for the spatial
components is similar.

\setcounter{equation}{0}
\section{Energy-momentum tensor from conservation laws}

The energy-momentum tensor $T_{\mu\nu}(x)$ of the collective excitation
 satisfies the following conservation law
\beq\label{cons1}
\del_\mu\,T^{\mu\nu}(x)\,=\,-F^{\nu\mu}_a(x)\,j_\mu^a(x),\eeq
for fields satisfying (\ref{ava}). This equation may be viewed as
the thermal average of the   conservation law formally
 satisfied by the corresponding field operators in the presence of
the external perturbation $j_\mu^a$. (Since no ultraviolet divergence
appears at the level of the present approximation\cite{Braaten90b,us},
we can safely ignore all the complications concerning the finiteness of
  $T_{\mu\nu}(x)$ in perturbation theory\cite{Callen}.)
Note, however, that
we shall not derive  $T_{\mu\nu}(x)$ by directly evaluating the
 expectation value of some local composite operator (e.g., the canonical
energy-momentum tensor associated to the QCD lagrangian),
but, rather, by integrating the conservation law (\ref{cons1}).
This is possible because, in the present approximation, all the
dynamical information is contained in the equation of motion (\ref{ava})
for the gauge fields, and we know, from eqs.~(\ref{solj}) or ~(\ref{j2}),
the explicit expression of the induced current in terms
of the  fields.

In order to construct the tensor $T^{\mu\nu}$, it is convenient to
 separate the standard Yang-Mills contribution, by writing
\beq\label{Ttheta}
T^{\mu\nu}(x)\equiv T_{YM}^{\mu\nu}(x) \,+\,\Theta^{\mu\nu}(x),\eeq
with \beq\label{T0}
T_{YM}^{\mu\nu}(x)\equiv -F_a^{\mu\rho}(x)\,F^\nu_{a\,\rho}(x)\,+\,
\frac{1}{4}\,g^{\mu\nu}\,F_a^{\alpha\beta}(x)\,F^a_{\alpha\beta}(x).\eeq
The tensor (\ref{T0}) satisfies the structural equation
\beq\label{dT0}
\del_\mu\,T_{YM}^{\mu\nu}(x)\,=\,-\,\left [\, D^\mu,\, F_{\mu\rho}(x)\,\right
]^a
\, F_a^{\nu\rho}(x).\eeq
In deriving this equation, the following identity is useful:
\beq\label{der}
\del_\mu\big (A^a\,B^a\big )\,=\,[D_\mu,\,A]^a\,B^a\,+\,
A^a\,[D_\mu,\,B]^a.\eeq
We use now the equations of motion (\ref{ava}) for $F^{\mu\nu}$
to transform eq.~(\ref{dT0}) into
\beq\label{DT0}
\del_\mu\,T_{YM}^{\mu\nu}(x)\,=\,-\, F_a^{\nu\rho}(x)\,(j^a_\rho(x)\,+\,
j^{ind\,a}_\rho(x)).\eeq
By comparing eqs.~(\ref{cons1}), (\ref{Ttheta}) and (\ref{DT0}),
we conclude that $\Theta^{\mu\nu}(x)$
 satisfies \beq\label{cons2}
\del_\mu\,\Theta^{\mu\nu}(x)\,=\,F^{\nu\mu}_a(x)\,j_\mu^{ind\,a}(x).\eeq
In the next sections we shall use either eq.~(\ref{j1}) or eq.~(\ref{j2})
for $j_\mu^{ind}$ in order to write  the r.h.s. of eq.~(\ref{cons2})
as a total derivative of some gauge-invariant  functional of the fields,
and  identify in this way  $\Theta^{\mu\nu}(x)$.

The tensor  $T_{\mu\nu}$ thus defined will  satisfy, by construction, the
conservation law (\ref{cons1}), and will vanish for large negative times.
Accordingly,  the 4-momentum constructed from it,
\beq\label{PT}
P^\nu(t)\equiv\int d^3 x\,T^{0\nu}(t,{\bf x}),\eeq
will coincide, at any time, with the correct 4-momentum of the collective
excitation, which is obtained by integrating the conservation law
\beq\label{Pcons}
\del_0\,P^\nu(t)\,=\,-\,\int d^3 x \,F^{\nu\mu}_a(x)\,j_\mu^a(x).\eeq
(In priciple, this equation determines $P^\nu(t)$, given the
initial condition $P^\nu(t\to -\infty)=0$. Recall
 that the external
sources and the average fields are assumed to vanish at $t\to -\infty$).
Of course,  the tensor   $T_{\mu\nu}$ thus obtained
 is not unique, since any two tensors which would differ
by a divergence-free piece, and lead to the same $P^\mu$, would be equally
acceptable. We shall indeed present in this section three different versions
of the energy-momentum tensor.

\subsection{A positive definite energy density}

Eq.~(\ref{j1}) for $j_\mu^{ind}$ allows us to write the r.h.s. of
eq.~(\ref{cons2}) as
\beq\label{rhs}
F^{\nu\mu}_a(x)\,j_\mu^{ind\,a}(x)&=&
3\,\omega^2_p\int\frac{d\Omega}{4\pi}\,F^{\nu\mu}_a(x)\,
v_\mu\,W_a^0(x;v)\nonumber\\
&=&3\,\omega^2_p\int\frac{d\Omega}{4\pi}\,
\left[ v\cdot D_x,\, W^\nu(x;v)\right]^a\,W_a^0(x;v),\eeq
where eq.~(\ref{eqw}) for $W^\nu(x;v)$ has been used in the second line.
For $\nu =0$, this can be easily written as a total derivative,
\beq\label{dth0}
{\bf E}^a(x)\cdot {\bf
j}_a^{\,ind}(x)&=&3\,\omega^2_p\int\frac{d\Omega}{4\pi}\,
\left[ v\cdot D_x,\, W^0(x;v)\right]^a\,W_a^0(x;v)\nonumber\\
&=&\del_\mu\,\Theta^{\mu 0}(x),\eeq
with the definition \beq\label{theen}
\Theta^{\mu 0}(x)\,\equiv\,\frac{3}{2}\,
\omega^2_p\int\frac{d\Omega}{4\pi}\,v^\mu\,W_a^0(x;v)\,W_a^0(x;v),\eeq
where the identity (\ref{der}) was used. After adding
$T_{YM}^{\mu 0}(x)$, eq.~(\ref{T0}), we end up with the
following expressions for the energy density
\beq\label{enden}
T^{00}(x)\,=\,\frac{1}{2}\Bigl({\bf E}^a(x)\cdot{\bf E}^a(x)\,+\,
{\bf B}^a(x)\cdot{\bf B}^a(x)\Bigr)\,+\,\frac{3}{2}\,
\omega^2_p\int\frac{d\Omega}{4\pi}\,W_a^0(x;v)\,W_a^0(x;v),\eeq
and for the energy flux density, or Poynting vector,
$S^i\equiv T^{i0}$,
\beq\label{Poyn}
{\bf S}(x)\,=\,{\bf E}^a(x)\times{\bf B}^a(x)\,+\,
\frac{3}{2}\,
\omega^2_p\int\frac{d\Omega}{4\pi}\,\,{\bf v}\,\,W_a^0(x;v)\,W_a^0(x;v).\eeq
The chromomagnetic field components are, as usual,
$B^i_a(x)\equiv -(1/2)\epsilon^{ijk}\,F^{jk}_a(x)$.

The expression (\ref{enden}) for $T^{00}(x)$ is manifestly positive
semidefinite and the same is  obviously true, at any time $t$,
for the excitation energy ${\cal E}(t)$,
\beq\label{entot}
{\cal E}(t)\equiv\int d^3 x\,T^{00}(t,\,{\bf x})\,=\,
{\cal E}_{YM}(t)\,+\,{\cal W}(t).\eeq
Note that
\beq\label{enloc}
{\cal E}_{YM}(t)\equiv \int d^3 x\,
\frac{1}{2}\left ({\bf E}^a(x)\cdot{\bf E}^a(x)\,+\,
{\bf B}^a(x)\cdot{\bf B}^a(x)\right)\eeq
also involves medium effects, since
 ${\bf E}^a(x)$ and ${\bf B}^a(x)$ are the {\it total} gauge fields,
including the plasma polarization. The second piece of (\ref{entot}),
\beq\label{ennon}
{\cal W}(t)\equiv \frac{3}{2}\,
\omega^2_p\int d^3 x\int\frac{d\Omega}{4\pi}\,\,W_a^0(x;v)\,W_a^0(x;v),\eeq
may be interpreted as the {\it polarization energy} of the plasma, that is,
the energy transferred by the gauge fields to the plasma constituents
 as mechanical work of the chromoelectric field. This becomes more obvious
if eq.~(\ref{dth0}) is used to reexpress ${\cal W}(t)$ as
\beq\label{work}
{\cal W}(t)\,=\,\int_{-\infty}^t dt^\prime \,\int d^3 x
\,{\bf E}^a(t^\prime,\,{\bf x})\cdot {\bf j}_a^{\,ind}(t^\prime,\,{\bf x}).\eeq
One can easily verify -- by explicitly performing the integral over
$t^\prime$ in the r.h.s. -- that eq.~(\ref{work}) leads indeed to the
 expression (\ref{ennon}) for ${\cal W}(t)$ if eq.~(\ref{j1})
for the induced current is used.

The positivity of the polarization energy (\ref{ennon}), and hence of the
total excitation energy (\ref{entot}),  reflects the stability of the
 equilibrium state  with respect to long wavelength  color fluctuations.

We briefly discuss now the other components of the energy-momentum tensor
 $\Theta^{\mu\nu}(x).$
We can write eq.~(\ref{cons2}) as
\beq\label{cons2a}
\del_\mu\,\Theta^{\mu\nu}(x)\,=\,\int\frac{d\Omega}{4\pi}\,
F^{\nu\mu}_a(x)\,{\cal J}_\mu^a(x;v),\eeq
where, according to eq.~(\ref{j1}),
${\cal J}^\mu_a(x;v)\equiv 3\,\omega^2_p\,v^\mu\,W_a^0(x;v)$.
We denote by $\Pi^\nu(x;v)$  the solution to
\beq\label{pi1}
(v\cdot\del_x)\,\Pi^\nu(x;v)\,=\,
F^{\nu\rho}_a(x)\,{\cal J}_\rho^a(x;v),\eeq
which vanishes for $x_0\to -\infty$, that is,
\beq\label{solpi1}
\Pi^\nu(x;v)\,=\,\int_0^\infty d\tau
F^{\nu\rho}_a(x-v\tau)\,{\cal J}_\rho^a(x-v\tau;v).\eeq
It is then easy to see that
\beq\label{themn1}
\Theta^{\mu\nu}(x)&=&\int\frac{d\Omega}{4\pi}\,v^\mu\,
\Pi^\nu(x;v)\eeq
satisfies eq.~(\ref{cons2a}).
For $\nu =0$, the integral over $\tau$ in (\ref{solpi1})
can be easily performed, with the result $\Pi^0(x;v)\,=\,
(3/2)\,\omega_p^2\,W^0_a(x;v)\,W^0_a(x;v)$.
 By inserting this result in eq.~(\ref
{themn1}), we recover our previous expression for $\Theta^{\mu 0}(x)$,
eq.~(\ref{theen}).
The physical content of the function $\Pi^\nu(x;v)$ is
transparent: it represents the density of the total
4-momentum transferred from the fields to
the hard particles with velocity ${\bf v}$ from time $\to -\infty$
 to $x_0$. In particular,  $\Theta^{0\nu}(x)$ is simply the angular
average of  $\Pi^\nu(x;v)$ over the possible orientations of ${\bf v}$.

The expression
(\ref{themn1})  for $\Theta^{\mu\nu}(x)$, though formally simple,
 involves generally a double integral along the hard particle trajectory
(see eqs.~(\ref{solpi1}) and (\ref{W})). We shall construct in
the following section a different expression for
the energy-momentum tensor, which  involves a single integral
along the trajectory.

\subsection{Another expression for $T^{\mu\nu}$}

We use now the second version for the induced color current
$j^{ind}_{\mu\,a}$, eq.~(\ref{j2}), in order to construct
another energy-momentum tensor,
$\tilde T^{\mu\nu}(x)\equiv T_{YM}^{\mu\nu}(x)+\tilde\Theta^{\mu\nu}(x)$,
different from the one obtained  in Sec.3.1, but physically equivalent.

When using eq.~(\ref{j2}) for  $j^{ind}_{\mu\,a}$, the r.h.s. of eq.~(\ref
{cons2}) reads
\beq\label{Fj}
F^{\nu\mu}_a(x)\,j_\mu^{ind\,a}(x)&=&
\frac{3}{2}\,\omega^2_p\int\frac{d\Omega}{4\pi}\left \{F^{\nu\mu}_a(x)\,
W_\mu^a(x;v)\,+\,F^{\nu\mu}_a(x)\, v_\mu\,\frac{\del W_a^\rho(x;v)}
{\del v^\rho}\right \}.\eeq
In the first term inside the brakets, we express $F^{\nu\mu}_a(x)$
  in terms of the functions $W^\mu_a(x;v)$, using the identity
\beq\label{WF}
 [D^\nu,\,W^{\mu}(x;v)]\,-\,[D^\mu,\,W^{\nu}(x;v)]\,=\,F^{\mu\nu}(x)\,+\,
ig\,[ W^{\mu}(x;v),W^{\nu}(x;v)].\eeq
(To prove this identity, act on eq.~(\ref{eqw}) with the covariant
derivative $D^\nu$. Then, consider the similar equation where
the indices $\mu$ and $\nu$  have been interchanged, and take the difference
of the two equations. The identity (\ref{WF}) follows after some repeated use
of the Jacobi identity.) We obtain \beq\label{FW1}
F^{\nu\mu}_a(x)\,W_\mu^a(x;v)\,=\,[D^\mu,\,W^{\nu}(x;v)]^a\,W_\mu^a(x;v)\,-
\,\frac{1}{2}\,\del^\nu\left (W\cdot W\right ),\eeq
with the notation $W\cdot W\equiv W_\mu^a(x;v)W^\mu_a(x;v)$.
For the second term in the r.h.s. of
(\ref{Fj}), we use eq.~(\ref{eqw}) for $W_\mu^a(x;v)$ to write
\beq\label{FW2}
F^{\nu\mu}_a\, v_\mu\,\frac{\del W_a^\rho}
{\del v^\rho}&=&\left[ v\cdot D_x,\, W^\nu\right]^a\,
\frac{\del W_a^\rho}{\del v^\rho}\nonumber\\&=&(v\cdot \del_x)
\left (W_a^\nu\,\frac{\del W_a^\rho}{\del v^\rho}\right )\,-\,
W_a^\nu \left[ v\cdot D_x,\,\frac{\del W^\rho}{\del v^\rho}
\right]^a.\eeq
By using eq.~(\ref{eqw}) once again, we see that
\beq \frac{\del}{\del v^\rho} \left[ v\cdot D_x,\, W^\rho(x;v)\right]^a\,=\,
 \frac{\del}{\del v^\rho} \left (F_a^{\rho\mu}(x)\,v_\mu\right )\,=\,0,\eeq
so that \beq\label{FW3}
F^{\nu\mu}_a\, v_\mu\,\frac{\del W_a^\rho}
{\del v^\rho}&=& \del_\mu
\left (v^\mu\,W_a^\nu\,\frac{\del W_a^\rho}{\del v^\rho}\right )\,+\,
W_a^\nu\,[ D^{\mu},\,W_{\mu}]^a.\eeq
By adding together eqs.~(\ref{FW1}) and (\ref{FW3}), we finally obtain
\beq
F^{\nu\mu}_a(x)\,j_\mu^{ind\,a}(x)&=&\del_\mu\,\tilde\Theta^{\mu\nu}(x),\eeq
with
\beq\label{tilt}
\tilde\Theta^{\mu\nu}(x)\equiv\frac{3}{2}\,\omega^2_p\int\frac{d\Omega}{4\pi}
\left \{W_a^\mu\,W_a^\nu\,+\,v^\mu\,W_a^\nu\,
\frac{\del W_a^\rho}{\del v^\rho}\,-\,\frac{1}{2}\,g^{\mu\nu}
\Bigl(W\cdot W\Bigr)\right \}.\eeq
The polarization
energy density associated to (\ref{tilt}), which is
\beq\label{tenden}
\tilde\Theta^{00}(x)&=&\frac{3}{2}\,\omega^2_p\int\frac{d\Omega}{4\pi}
\left \{W_a^0\,W_a^0
\,+\,W_a^0\,\frac{\del W_a^\rho}{\del v^\rho}\,-\,\frac{1}{2}\,
\Bigl(W\cdot W\Bigr)\right \},\eeq
is different from the previous expression in Sec. 4, eq.~(\ref{theen}).
In particular, the expression above is {\it not} manifestly
positive.
One advantage of the new tensor above is that eq.~(\ref{tilt})
offers, for the  components $(\mu,i)$,
  a simpler non-local structure then eq.~(\ref{themn1}).

It is possible to show that the energy-momentum tensor
$\tilde T^{\mu\nu}(x)$ is physically equivalent
to that proposed by Weldon in Ref.\cite{Weldon92}.
Indeed, the difference between the two tensors is the total derivative
of an antisymmetric tensor of rank 3. To be specific, we have
$\tilde T^{\mu\nu}(x)-T_W^{\mu\nu}(x)\,=\,\del_\rho\,X^{\rho\mu\nu}(x)$, with
\beq\label{usvsW}
X^{\rho\mu\nu}(x)\,\equiv\, \frac{3}{2}\,\omega^2_p\int\frac{d\Omega}{4\pi}
   \Bigl\{\Bigl(
v^\rho\,{\cal V}^\mu_a(x;v)\,-\,v^\mu\,{\cal V}^\rho_a(x;v)\Bigr)
W_a^\nu(x;v)\Bigr\},\eeq
and $X^{\rho\mu\nu}(x)\,=\,-\,X^{\mu\rho\nu}(x)$. Here,
 $T_W^{\mu\nu}(x)$ denotes  Weldon's tensor and
the functions ${\cal V}^\mu_a(x;v)$ satisfy, by definition, the
following equation:
\beq
\left[ v\cdot D_x,\, {\cal V}^\mu(x;v)\right]^a\,=\,W_a^\mu(x;v).\eeq
Eq.~(\ref{usvsW}) is obtained by first transforming  Weldon's  original
expression with the help of the identity (\ref{WF}).

The tensors $\Theta^{\mu\nu}$ and $\tilde\Theta^{\mu\nu}$ constructed up to
now are not symmetric, which is not surprising since
the indices $\mu$ and $\nu$ enter disymmetrically  in
 the conservation law (\ref{cons2}), on which
our derivation is based.

\subsection{A traceless and symmetric tensor}

We construct now a symmetric tensor,
 starting again from eq.~(\ref{cons2}), but
using in its r.h.s.  the expression (\ref{jind}) for the induced
current, which we write as
\beq\label{js}
j_\mu^{ind\,a}(x)\,=\,g\int\frac{d^3k}{(2\pi)^3}\,\frac{k_\mu}{k}
\,\delta f^a({\bf k},x),\eeq
with
\beq\label{df}
\delta f^a({\bf k},x)\,\equiv\,N_{\rm f}\left[ \delta
 n_+^a({\bf k},x)-\delta n_-^a({\bf k},x)\right]\,+\,
2N\,\delta N^a({\bf k},x).\eeq
The r.h.s. of eq.~(\ref{cons2}) reads then
\beq\label{rhs3}
F^{\nu\mu}_a(x)\,j_\mu^{ind\,a}(x)&=&
g\int\frac{d^3k}{(2\pi)^3}\,F^{\nu\mu}_a(x)\,v_\mu\,\delta f^a({\bf k},x).\eeq
We use  the simple identity
\beq
F^{\nu\mu}_a(x)\,v_\mu\,=\,F^{i\mu}_a(x)\,\frac{\del}{\del k^i}
\,(v^\nu\,k_\mu),\eeq
(recall that $v^\nu = k^\nu/k^0$ and $k^0=|{\bf k}|$),
and an integration by parts, to rewrite
(\ref{rhs3}) as
\beq\label{rhs4}
F^{\nu\mu}_a(x)\,j_\mu^{ind\,a}(x)&=&-\,
g\int\frac{d^3k}{(2\pi)^3}\,v^\nu\,F^{i\mu}_a(x)\,k_\mu\,\frac{\del}{\del k^i}
\,\delta f^a({\bf k},x).\eeq
 Let $\delta\epsilon ({\bf k}, x)$ be the solution to
\beq\label{deps}
(v\cdot\del_x)\,\delta\epsilon ({\bf k}, x)\,=\,-g\,F^{i\mu}_a(x)\,k_\mu\,\frac
{\del}{\del k^i}\,\delta f^a({\bf k},x),\eeq
which vanishes when $x_0\to -\infty$. Then
\beq \delta\epsilon ({\bf k}, x)\,=\,
N_{\rm f}\,\Bigl[\delta\epsilon_+ ({\bf k}, x)\,+\,\delta\epsilon_-
 ({\bf k}, x)\Bigr]\,+\,2N\,\delta\epsilon_{\rm g}({\bf k}, x),\eeq
with obvious notations, and eq.~(\ref{rhs4}) becomes
\beq F^{\nu\mu}_a(x)\,j_\mu^{ind\,a}(x)&=&\del_\mu\,
\Theta_{{\rm sym}}^{\mu\nu}(x),\eeq
where \beq\label{thetas}
\Theta_{{\rm sym}}^{\mu\nu}(x)\,\equiv\,
\int\frac{d^3k}{(2\pi)^3}\,v^\mu\,v^\nu\Bigl\{
N_{\rm f}\,[\delta\epsilon_+ ({\bf k}, x)\,+\,\delta\epsilon_-
 ({\bf k}, x)]\,+\,2N\,\delta\epsilon_{\rm g}({\bf k}, x)\Bigr\}\eeq
is a {\it traceless symmetric} tensor. The same is therefore true
 for the corresponding
 energy-momentum tensor $T_{{\rm sym}}^{\mu\nu}(x)\equiv
T_{YM}^{\mu\nu}(x)\,+\,\Theta_{{\rm sym}}^{\mu\nu}(x)$.
Remark that, in the
weak field limit,  $\delta\epsilon ({\bf k}, x)$ is quadratic
 with respect to the gauge fields (see eq.~(\ref{deps})), as opposed to
 $\delta f^a ({\bf k}, x)$ which, in the same limit, is linear in
$A_\mu$.

The symmetry of $T_{{\rm sym}}^{\mu\nu}(x)$ allows us to construct a
simple density for the generalized angular momentum of the  gauge fields:
\beq\label{ang}
M^{\lambda\mu\nu}(x)\equiv x^\mu\,T_{{\rm sym}}^{\lambda\nu}(x)\,-\,
x^\nu\,T_{{\rm sym}}^{\lambda\mu}(x).\eeq
It can be verified, by using eq.~(\ref{cons1}), that
$M^{\lambda\mu\nu}(x)$ satisfies the correct conservation law \beq
\del_\lambda\,M^{\lambda\mu\nu}(x)\,=\,-\left (x^\mu\,F_a^{\nu\lambda}(x)
\,-\,x^\nu\,F_a^{\mu\lambda}(x)\right )\,j^a_\lambda(x),\eeq
where one recognizes in the r.h.s. the torque applied to the system
by the external current. Since $M^{\lambda\mu\nu}(x)$
 vanishes when $x_0\to -\infty$, the total angular momentum is
 correctly given by
\beq\label{tang}
J^{\mu\nu}(t)\,=\,\int d^3 x\,M^{0\mu\nu}(x),\eeq
for any $t$.

The vanishing trace of  $T_{{\rm sym}}^{\mu\nu}(x)$ reflects
 the dilatation invariance of the  massless tree-level QCD lagrangian.
At the level of the present approximation, there is no breaking of the
 dilatation symmetry, neither by radiative corrections (via the
regularization procedure), nor by a  thermally induced mass, which
is negligible for the hard particles.

\setcounter{equation}{0}
\section{Applications}

In this section we use the general expressions   derived in Sec.3
to study  the energy transfer
in the plasma and the energy-momentum content of particular excited states.
We   consider mostly the weak field, or abelian, limit, where the
components of $T^{\mu\nu}$ are quadratic
in the gauge potentials. We show that,
in this particular limit, our expressions are the straightforward
generalization of the corresponding  formulae in the  standard
plasma physics.

Moreover, any solution of the abelian gauge field and  kinetic
equations can be directly imbedded in  the corresponding
non-abelian equations.
That is, let $A_\mu(x)$ and $W^0(x;v)$ be solutions to the coupled abelian
equations
\beq\label{aba}
\del^\nu F_{\nu\mu}(x)\,-\, 3\,\omega^2_p\int\frac{d\Omega}{4\pi}
\,v_\mu\,W^0(x;v)&=&j_\mu(x),\eeq
and
\beq\label{abw}
(v\cdot\del_x)\, W^0(x;v)&=&{\bf v}\cdot{\bf E}(x).\eeq
Then, $A_\mu^a(x)\equiv \lambda^a\,A_\mu(x)$ and $W_a^0(x;v)\equiv
\lambda^a W^0(x;v)$, with constant $\lambda^a$, satisfy
the corresponding non-abelian equations, (\ref{ava}) and (\ref{eqw}),
for an external  source pointing in a fixed direction in  color
space, $j_\mu^a(x)=\lambda^a j_\mu(x)$.
This property extends the corresponding one for classical Yang-Mills
equations in vacuum (see, e.g., Ref. \cite{Actor}).

A slightly more general field configuration for which the general solution
is still trivial corresponds to gauge fields
in the  Cartan subalgebra of $SU(N)$, that is,
$A^i(t)\equiv \sum_s A^i_s(t) T^s$,
where the indice $s$ takes on $N-1$ values corresponding to the $N-1$
commuting generators $T^s$ of $SU(N)$.  Then, clearly, $F^{\mu\nu}_s=
\del^\mu A^\nu_s-\del^\nu A^\mu_s$, and  $F^{\mu\nu}_a=W_a^\mu=j_a^\mu
=0$ for $a\not = s$. The non-trivial components $A^\mu_s$, $W^\mu_s$
and $j^\mu_s$ satisfy abelian-like equations, as eqs.~(\ref{aba})--(\ref{abw})
 above. Hence, the problem reduces itself to $N-1$ independent abelian
theories.

 In subsection 4.1, we briefly discuss the case of static gauge fields.
In the following subsection, we  investigate the energy transfer
between  fields and particles and, in particular, the energy dissipation
through the Landau mechanism.  Finally, in subsection 4.3, we consider
 some interesting special field configurations,
namely, time-periodic fields and abelian plasma waves.

\subsection{Static  field configurations}

We consider a field configuration
 induced by a static external color density
 $\rho^a({\bf x})$. We choose a gauge where the
potentials $A_\mu^a$  are time-independent.
In this case, it is useful to reexpress $W_a^0(x;v)$,
eq.~(\ref{W}), as a function of the gauge potentials themselves
\beq\label{Wt}
W_a^0(x;v)&=&-A_0^a(x)\,+\,\int _0^\infty du\,U_{ab} (x,x-vu)\,v\cdot
{\dot A}^b(x-vu),\eeq
where ${\dot A}^a(x)\equiv\del_0 A^a(x)$. For a static field configuration,
${\dot A}^a(x)=0$ and
\beq
W_a^0({\bf x};v)&=&-A^0_a({\bf x}).\eeq
It follows that the induced current (\ref{j1}) reduces to a static
color  distribution
\beq\label{stj}
j^{ind}_{\mu\,a}({\bf x})=-\,g_{\mu 0}\,3\,\omega_p^2\,A^0_a({\bf x})
\equiv -\,g_{\mu 0}\,\rho_a^{ind}(x).\eeq
Then, the gauge field equations (\ref{ava}) become simply
\beq\label{snon}
[D_i, E^i({\bf x})]\,+\,3\,\omega_p^2\,A^0({\bf x})\,=\,\rho({\bf x}),
\nonumber\\ \epsilon^{ijk}[D_j, B^k]\,=\,i\,g\,[A^0, E^i].\eeq
The first equation above clearly shows the Debye screening of the static
chromoelectric field, with the screening length
$\lambda_D\,=\,1/(3\,\omega_p^2)^{1/2}$.
Note also that the chromomagnetic field induced by a static color
charge is generally  not zero in a  non-abelian theory.
Furthermore,
\beq\label{stthe}
\Theta^{00}({\bf x})&=&\frac{3}{2}\,\omega_p^2\, A_a^0({\bf x})\,A_a^0({\bf x})
\,=\,-\frac{1}{2}\,\rho_a^{ind}({\bf x})\,A^0_a({\bf x}).\eeq
In analogy with what is commonly done in abelian plasmas, (see Appendix),
we define the {\it color polarization vector}
 ${\bf P}^a({\bf x})$ by $[ D^i, P^i({\bf x})]^a\equiv \rho_a^{ind}({\bf x})$,
and construct the {\it color displacement vector} ${\cal {\bf D}}^a(x)\equiv
{\bf E}^a(x)+{\bf P}^a(x)$.
 Then,  the polarization energy (\ref{ennon}) can be written as
\beq
{\cal W}\,=\,\int d^3 x \,\Theta^{00}({\bf x})
\,=\,-\frac{1}{2}\int d^3 x\, [ D^i, P^i({\bf x})]^a\,A_0^a({\bf x})
\,=\,\frac{1}{2}\int d^3 x\, {\bf E}^a({\bf x})\cdot {{\bf P}}^a({\bf x}),\eeq
as familiar in electrostatics of classical dielectric media\cite{Landau}.
(We used here $E^i_a({\bf x})=[D^i,\,A^0({\bf x})]^a$ for static fields,
together with eq.~(\ref{der}) and an integration by parts.) The total field
energy follows:
\beq\label{sten}{\cal E}&=&\frac{1}{2}\int d^3 x\,\Bigl({\bf E}^a({\bf x})
\cdot {\cal {\bf D}}^a({\bf x})\,+\,{\bf B}^a({\bf x})\cdot
{\bf B}^a({\bf x})\Bigr).\eeq

For an abelian plasma, we can solve the equation
\beq
(-\Delta\,+\,3\,\omega_p^2)\,A^0({\bf x})&=&\rho({\bf x})\eeq
to get the gauge field
\beq
A^0({\bf x})&=&\int \frac{d^3 y}{4\pi}\,\frac{\rho ({\bf y})}{R}
\,e^{-R/\lambda_D},\eeq
where $R\equiv |{\bf x}-{\bf y}|$.
Then, the  field energy  can be expressed in terms
of the external charge, with the standard result
\beq {\cal E}&=&\frac{1}{2}\int d^3 x\,\left ({\bf E}^2({\bf x})\,+\,
3\,\omega_p^2\,A_0^2({\bf x})\right )\,=\,
\frac{1}{2}\int\frac{d^3 p}{(2\pi)^3}\,\frac{\rho({{\bf p}})\,\rho(-{{\bf p}})}
{p^2\,+\,3\,\omega_p^2}.\eeq

For a non-abelian plasma, the non-linear structure of the generalized Maxwell
equations  (\ref{snon}) does not allow such a simple elimination of
 the gauge fields in terms of $\rho^a$.

\subsection{Energy transfer and Landau damping}

When studying time-dependent field configurations, there is a new feature
 arising: the imaginary part of the polarization tensor (\ref{P}) is
non-vanishing, which indicates that there is dissipation of the gauge
field energy in the medium. To investigate this,  both for abelian and
non-abelian fields, we  compute
  the rate of energy absorbtion by the plasma constituents (see
eq.~(\ref{work})):
\beq\label{ratew}
\frac{d\,{\cal W}(t)}{d\,t}\,=\,\int d^3 x
\,{\bf E}^a(t,\,{\bf x})\cdot {\bf j}_a^{\,ind}(t,\,{\bf x}).\eeq

Consider first an  abelian-like plasma, in
 which case the general description
in Sec.2 reduces  to the standard Maxwell equation in a polarizable
medium\cite{PhysKin}.
For simplicity, consider a  time-periodic field
\beq\label{qper} {\bf E}
(t,\,{{\bf p}})\,=\,{\bf E}({{\bf p}})\,e^{\eta t}\,
\cos \omega_0 t,\eeq
with $\eta\to 0_+$ and both $\omega_0$ and $p\equiv |{{\bf p}}|$
of order $gT$.
Then, a simple calculation shows that the average energy loss of the field,
that is, the expectation value of eq.~(\ref{ratew}) over the period
$T_0=2\pi/\omega_0$, is   proportional to the imaginary part of
the {\it dielectric tensor}\cite{PhysKin}:
\beq\label{erate}
\left\langle \frac{d\,{\cal W}}{d\,t}\right\rangle &=&\frac{1}{2}
\int \frac{d^3 p}{(2\pi)^3}\,\,\omega_0\,E^i(-{{\bf p}})\left (
{\rm Im}\,\epsilon ^{ij}(\omega_0,\,{{\bf p}})\right )\,E^j({{\bf p}}).\eeq
(The definition of  $\epsilon^{ij}$, as well as its relation with the
polarization tensor, are reviewed in Appendix).
In our case, eqs.~(\ref{eps}) and (\ref{P}) show that
\beq\label{imeps}
{\rm Im}\,\epsilon ^{ij}(\omega,\,{{\bf p}})\,=\,3
\,\pi\,\frac{\omega_p^2}{\omega}\int
\frac {d\Omega}{4\pi}\,\,v^i\,v^j\,\delta (\omega - {\bf v}\cdot{{\bf p}}),\eeq
and we obtain  the average energy loss as
\beq\label{avrate}
\left\langle
\frac{d\,{\cal W}}{d\,t}\right\rangle &=&\frac{3}{2}
\,\pi\,\omega_p^2\int \frac{d^3 p}{(2\pi)^3}\int
\frac {d\Omega}{4\pi}\,\,\delta (\omega_0 - {\bf v}\cdot{{\bf p}})
\,\Big |{\bf v}\cdot{\bf E}({{\bf p}})\Big |^2.\eeq
This expression is obviously positive:
  on the average, the fields transfer energy to the particles.
This energy dissipation is related to the {\it Landau damping} of the mean
fields\cite{PhysKin}, that
is, to the coherent transfer of energy in between the gauge fields and
the hard particles which are moving in phase with the field
oscillations.
As the $\delta$ function clearly shows, the dissipation
is kinematically allowed only for fields carying
space-like momenta (i.e., when $\omega_0={\bf v} \cdot {\bf p}$).

The energy given by the fields to the colored particles is ultimately
supplied by the external source, that is
\beq\left\langle \frac{d\,{\cal W}}{d\,t}\right\rangle &=&-
\left\langle\int d^3 x\,{\bf E}(x)\cdot {\bf j}(x)\right\rangle ,\eeq
as shown by the conservation law (\ref{cons1}) (the local piece
$d{\cal E}_{YM}/dt$ averages to 0). Therefore, in order to compute the
energy loss of a given external source, one just has to replace the
electric field $E^i$ in eq.~(\ref{erate}) in terms of the current $j^\mu$,
by solving the Maxwell equation (\ref{ava}). In the abelian regime, this
can be easily done by Fourier analysis,  and leads to the following
expression for the average energy loss of the time-periodic external current
 ${\bf j}({{\bf p}})\,\cos \omega_0 t$:
\beq\label{jrate}
\left\langle \frac{d\,{\cal W}}{d\,t}\right\rangle =-\frac{1}{2}
\int \frac{d^3 p}{(2\pi)^3}\,\omega_0
\left\{\frac{|\rho({{\bf p}})|^2}{p^2}\,{\rm Im}\,\frac{1}{\epsilon_l
(\omega_0,p)} +\frac{|{\bf j}_t({{\bf p}})|^2}{\omega_0^2}\,
{\rm Im}\,\frac{1}{\epsilon_t
(\omega_0,p)-{p^2}/{\omega_0^2}}\right\}.\,\,\eeq
Here,  $\omega_0 \,\rho({{\bf p}})=-i{{\bf p}}\cdot {\bf j}({{\bf p}})$,
as required by the current conservation, while
${\bf j}_t({\bf p})$ is the transverse component of ${\bf j}({{\bf p}})$,
i.e., $j_t^i({\bf p})=(\delta^{ik}- \hat p^i \hat p^k)\,j^k({\bf p})$.
The  longitudinal and transverse pieces of
 $\epsilon ^{ij}$ are defined in Appendix (eqs.~(\ref{elt})--(\ref{et})).

Clearly, eq.~(\ref{jrate})  also holds for non-abelian field configurations
such that
the space-time and the color degrees of freedom decouple, that is, for
the imbedded abelian solutions to eq.~(\ref{ava}) with a color source
 ${\bf j}^a(t,{{\bf p}})= \lambda^a {\bf j}(t,{{\bf p}})$, and
constant $\lambda^a$. (The only modification consists in a supplimentary
 factor $\lambda^a\lambda^a$ in the previous equation.)
 An expression similar to (\ref{jrate})
has been used in Refs. \cite
{Thoma91,Matsui91} in order to compute the energy loss in the plasma
of an external parton, which is assimilated to a classical point-like
colored particle, moving with constant velocity ${\bf v}$, and therefore
giving rise to a current
\beq\label{jpart}
{\bf j}^a
(\omega,\,{{\bf p}})\,=\,2\pi\,\lambda^a\,{\bf v}\,\delta (\omega-{\bf v}\cdot
{{\bf p}}).\eeq
Clearly, the energy loss calculated according to eq.~(\ref{jrate})
accounts only for the energy transferred by the external source
to the hard particles, through its coupling to the collective modes.
This is indeed the dominant process for long wavelength and low frequency
excitations.
However, these conditions are not satisfied
for the whole range of momenta involved in the  source (\ref{jpart}).
Strictly speaking, when used with such external sources,
all our previous integrals over ${{\bf p}}$ should
involve an ultraviolet cut-off of order $gT$.

To close this subsection, we return to the general case of an arbitrary
non-abelian gauge field configuration (with $|A_\mu^a| \simle T$)
and compute the total energy absorbed by the plasma
during the life-time of the excited state:
\beq\label{gloss}
\Delta {\cal W}\,\equiv\,\int_{-\infty}^\infty dt\,
\frac{d\,{\cal W}}{d\,t}\,=\,\int_{-\infty}^\infty dt\int d^3 x
\,{\bf E}^a(t,\,{\bf x})\cdot {\bf j}_a^{\,ind}(t,\,{\bf x}).\eeq
This quantity is well defined if we assume that
the external perturbation is switched off
 adiabatically at $t\to \infty$.
It coincides, of course, with the $t\to \infty$ limit of the polarization
energy ${\cal W}(t)$  given by eq.~(\ref{work}).
By using eq.~(\ref{solj}) for ${\bf j}_a^{\,ind}$, we obtain,
after some elementary changes in the integration variables,
\beq\label{Gloss}
\Delta {\cal W}
&=&\frac{3}{2}\,\omega^2_p\int\frac{d\Omega}{4\pi}
\int d^4 x\,{\bf v}\cdot{\bf E}^a(x)\,\int_{-\infty}^\infty du
\,U_{ab}(x,x-vu)\, {\bf v}\cdot{\bf E}^b(x-vu).\eeq
 This quantity vanishes if
\beq\label{nodamp}
\int_{-\infty}^\infty du
\,U_{ab}(x,x-vu)\, E_b^i(x-vu)\,=\,0,\eeq
for arbitrary $x$, ${\bf v}$ and $i$.
For fixed ${\bf v}$, we may choose the axial gauge $v^\mu A_\mu^a=0$. Then
eq.~(\ref{nodamp}) becomes
\beq\label{n1}
0\,=\,\int_{-\infty}^\infty du\,  E_a^i(x-vu)\,=\,\int\frac{d^4p}{(2\pi)^4}
\,e^{-ip\cdot x}\,\delta (p^0-{\bf v}\cdot {{\bf p}})\,E_a^i(p^0,{{\bf p}}).
\eeq
As it should be valid for any  $x$, eq.~(\ref{n1}) implies
$E_a^i(p^0={\bf v}\cdot {{\bf p}},{{\bf p}})=0$, which is the
condition for absence of Landau damping.
More generally, eq.~(\ref{nodamp}) characterizes the non-abelian
gauge fields which propagate without dissipation.
It is one of the conditions invoked
in Ref.\cite{us} in order to write a well-defined effective action for
the soft gauge fields.

\subsection{Plasma waves}

We start by reconsidering the periodic field  (\ref{qper}) and assume that
${\bf E}({{\bf p}})=0$ if $|{{\bf p}}|>\omega_0$,
so that there is no dissipation.
The average, over the period $T_0$, of the polarization energy
${\cal W}(t)$, eq.~(\ref{ennon}),  is
\beq\label{avpol}
\left\langle
{\cal W}\right\rangle &=&\frac{3}{4}
\,\pi\,\omega_p^2\int \frac{d^3 p}{(2\pi)^3}\int
\frac {d\Omega}{4\pi}\,\,\frac {\Big |{\bf v}\cdot{\bf E}({{\bf p}})\Big |^2}
{(\omega_0 - {\bf v}\cdot{{\bf p}})^2}.\eeq
The angular integral can be easily evaluated (see  eqs.~(\ref{vivj2}) and
 (\ref{ab})).
After adding the Yang-Mills contribution,
\beq\label{avloc}
\left \langle {\cal E}_{YM}\right\rangle\,=\,
\frac{1}{4}\int \frac{d^3 p}{(2\pi)^3}\,\left
\{\Big | {\bf E}({{\bf p}})\Big |^2\,+\,\frac {1}{\omega_0^2}
\,\Big |{{\bf p}}\times {\bf E}({{\bf p}})\Big |^2\right \},\eeq
we derive the following expression for the
averaged value (over $T_0$) of the total field energy
 ${\cal E}(t)={\cal E}_{YM}(t)+{\cal W}(t)$:
\beq\label{aven}
\left \langle {\cal E}\right\rangle =
\frac{1}{4}\int \frac{d^3 p}{(2\pi)^3}\left
\{ \left [1+3\,\frac{\omega_p^2}{p^2}\,a(\omega_0/p)\right ]
\Big | {\bf E}_l({{\bf p}})\Big |^2+\left [1+\frac{p^2}{\omega_0^2}+
3\,\frac{\omega_p^2}{p^2}\,b(\omega_0/p)\right ]
\Big | {\bf E}_t({{\bf p}})\Big |^2\right \}.\,\,\,\,\,\eeq
The coefficients $a(\omega_0/p)$ and $b(\omega_0/p)$ are given in
eq.~(\ref{ab}). They are both positive for $\omega_0/p > 1$.
Furthermore,  ${\bf E}({{\bf p}})\equiv {\bf E}_l({{\bf p}}) +
{\bf E}_t({{\bf p}})$, with  ${{\bf p}}\cdot {\bf E}_t({{\bf p}})=0$.

The coefficients in eq.~(\ref{aven}) can be related to the derivatives
of the dielectric tensor (\ref{eps}) with respect to $\omega$. Indeed,
by using eqs.~(\ref{el})--(\ref{et}) for $\epsilon_{l,\,t}(\omega, p)$,
it is straightforward to verify that eq.~(\ref{aven}) coincides with
\beq\label{Aven}
\left \langle {\cal E}\right\rangle =
\frac{1}{4}\int \frac{d^3 p}{(2\pi)^3}\left
\{ \frac {d}{d \omega_0}\Bigl[\omega_0\epsilon_l (\omega_0,p)\Bigr]
\Big | {\bf E}_l({{\bf p}})\Big |^2+\frac{d}{d\omega_0}\Bigl[
\omega_0(\epsilon_t(\omega_0,p)-p^2/\omega_0^2) \Bigr]
\Big | {\bf E}_t({{\bf p}})\Big |^2\right \}.\,\,\,\,\,\eeq
A similar expression holds  for ordinary polarizable
media\cite{Landau}.

The previous discussion can be easily generalized to study
the {\it abelian plasma waves}, i.e., the solutions to
eq.~(\ref{aba}) without external sources. The decomposition of an
arbitrary electric field in terms of normal modes is of the form
\beq\label{pwave}
{\bf E}(t,\,{\bf x})\,=\,\int \frac{d^3 p}{(2\pi)^3}\,e^{\,i{{\bf p}}\cdot {\bf
x}}
\,\sum_{s=l,t}\left \{ {\bf e}_s({{\bf p}})\,e^{-i\omega_s(p) t}\,+\,
 {\bf e}_s^{\,*}(-{{\bf p}})\,e^{i\omega_s(p) t}\right \},\eeq
where the frequencies $\omega_s(p)$ are solutions of the dispersion equations
\beq\label{disp}
\epsilon_{l}(\omega_l(p), p)\,=0,\qquad\,{\rm and}\,\qquad
\epsilon_{t}(\omega_t(p), p)\,=\,p^2/\omega^2,\eeq
for longitudinal and transverse modes,
respectively\cite{Klimov81,Weldon82,Pisarski89b}.
Note that  $\omega_l(0)=\omega_t(0)=\omega_p$
and  $\omega_{l,\,t}(p)>p$, for any $p$,
and hence the normal modes propagate without damping.

By using the dispersion relations (\ref{disp}), and
 eqs.~(\ref{lres})--(\ref{tres}), we easily obtain the
energy of the field (\ref{pwave}) as
\beq\label{ewave}
 {\cal E}&=&\int \frac{d^3 p}{(2\pi)^3} \Biggl\{
 \left (\frac{3\omega_p^2}{\omega_l^2(p)-p^2}-1\right )
{\bf e}_l({{\bf p}})\cdot {\bf e}_l^{\,*}({{\bf p}})\nonumber\\
&&\qquad+\,\left (
 \frac{3\omega_p^2}{\omega_t^2(p)-p^2}+\frac{p^2}{\omega_t^2
(p)}-1\right )
{\bf e}_t({{\bf p}})\cdot {\bf e}_t^{\,*}({{\bf p}})\Biggr\}.\,\,\eeq
As expected, this is time-independent.
The coefficients in eq.~(\ref{ewave}) are inversely proportional
to the residues  $z_{l,\,t}(p)$ of the effective  propagators (\ref{effd})
at the appropriate poles (see eqs.~(\ref{zlt}) and (\ref{eres})).
Since the energy (\ref{ewave}) is positive, the same holds for
 the residues $z_{l,\,t}(p)$; this  can be also   verified
directly by using the  formulae given in Appendix.

It is quite straightforward to verify that the second
expression for the energy density, i.e. eq.~(\ref{tenden}), leads
to the same excitation energy. More interestingly,
 eq.~(\ref{tilt}) allows us to
 compute easily the total  3-momentum of the field configuration (\ref{pwave}).
We obtain:
\beq\label{Pwave}
{\bf P}&=&\int \frac{d^3 p}{(2\pi)^3}\Biggl\{
 \frac{{{\bf p}}}{\omega_l(p)}
\left (\frac{3\omega_p^2}{\omega_l^2(p)-p^2}-1\right )
{\bf e}_l({{\bf p}})\cdot {\bf e}_l^{\,*}({{\bf p}})\nonumber\\
&&\qquad +\, \frac{{{\bf p}}}{\omega_t(p)}\left (
 \frac{3\omega_p^2}{\omega_t^2(p)-p^2}+\frac{p^2}{\omega_t^2
(p)}-1\right )
{\bf e}_t({{\bf p}})\cdot {\bf e}_t^{\,*}({{\bf p}})\Biggr\}.\eeq
We have used here eq.~(\ref{vivj3})
in  Appendix, as well as
the dispersion relations for the normal modes, eqs.~(\ref{disp}).

Physically more transparent expressions for the energy and the momentum
of the plasma waves are obtained by expressing them in terms of the gauge
potentials which correspond to the field (\ref{pwave}). We use the normal
modes decomposition
\beq\label{Awave}
A_\mu(x)\,=\,\int \frac{d^3 p}{(2\pi)^3}\sum_{\lambda=0}^2
\frac{z_\lambda^{1/2}(p)}{\sqrt {2\omega_\lambda(p)}}  \left \{
\epsilon_\mu({{\bf p}};\lambda)\,a_\lambda({{\bf p}})
\,e^{\,-ip\cdot x}\,+\,
\epsilon_\mu^*\,({{\bf p}};\lambda)\,a_\lambda^*\,({{\bf p}})
\,e^{\,ip\cdot x}\right \},\,\,\eeq
where $\lambda =0$ corresponds to the longitudinal mode,
$\lambda=1,\,2$ to the transverse ones,
$p^0=\omega_\lambda(p)$, and the polarization vectors
$\epsilon^\mu({{\bf p}};\lambda)$ are defined in Appendix.
In terms of $a_\lambda({{\bf p}})$ and $a_\lambda^*({{\bf p}})$,
the plasma wave energy and momentum become simply
 \beq\label{Aen}
{\cal E}&=&\int \frac{d^3 p}{(2\pi)^3}\left \{\omega_l(p)\,
a_l({{\bf p}})\,a_l^*(\,{{\bf p}})\,+\,\omega_t(p)\sum_{\lambda=1,2}
a_\lambda({{\bf p}})\,a_\lambda^*(\,{{\bf p}})\right \},\eeq
and, respectively,
\beq\label{AP}
{{\bf P}}&=&\int \frac{d^3 p}{(2\pi)^3}\left \{{{\bf p}}\,
a_l({{\bf p}})\,a_l^*(\,{{\bf p}})\,+\,{{\bf p}}\sum_{\lambda=1,2}
a_\lambda({{\bf p}})\,a_\lambda^*(\,{{\bf p}})\right \}.\eeq
The interpretation of the two equations above in terms of the
color elementary oscillations is  obvious.

\setcounter{equation}{0}
\section{Global color oscillations}

In this section we study particular non-abelian solutions to the field
 equations  which are uniform in  space, i.e., which describe global
color oscillations of the plasma.
For convenience, we choose the temporal gauge, $A^0_a(x)\,=\,0$. Therefore,
$A^\mu_a(x)\equiv (0, {\bf A}^a(t))$. For such fields, it is useful to
work with the following representation for  $W_a^0(x;v)$:
\beq
W_a^0(x;v)&=&-\,{\bf v}\cdot {\bf A}^a(x)\,-\,\int _0^\infty du\,U_{ab}
 (x,x-vu)\,\bigl({\bf v}\cdot{\bf \nabla}_x\bigr)\Bigl(v\cdot A^b(x-vu)
\Bigr).\eeq
This can be derived  from eq.~(\ref{Wt}), by noting that
 $U(x,x-vu)$ satisfies the equation
\beq \frac{\del}{\del u}\,U(x,x-vu)&=&-\,i\,g\,U(x,x-vu)\,v\cdot A(x-vu),
\eeq where $A^\mu\equiv A^\mu_a\, T^a$.  For fields which do not depend
on spatial coordinates, we have simply
\beq
W_a^0({t};v)&=&-\,{\bf v}\cdot {\bf A}^a({t}).\eeq
{}From eq.~(\ref{j1}), one obtains then
$\rho^{ind}_a(t)\,=\,0$ and ${\bf j}^{ind}_a(t)\,=\,-\omega_p^2\,{\bf A}^a(t)$,
and, using the definition (\ref{theen}),
\beq\label{thetat}
\Theta^{00}({t})&=&\frac{1}{2}\,\omega_p^2\,{\bf A}^a(t)\cdot {\bf A}^a(t).\eeq
This expression of $\Theta^{00}$ is very similar to that obtained
in the static case,
eq.~(\ref{stthe}); in particular, both eqs.~(\ref{stthe}) and (\ref{thetat})
are  local functionals of the fields.
The color electric and magnetic fields are
\beq\label{EBt}
{\bf E}^a(t)\,=\,-\,\frac{d{\bf A}^a}{dt}\qquad
{\bf B}^a(t)\,=\,\frac{g}{2}\,f^{abc}\,{\bf A}^b(t)\times {\bf A}^c(t).\eeq
Using these expressions, we obtain a simple expression for
the total energy density in terms of the vector potentials
\beq\label{ent}
T^{00}(t)&=&\frac{1}{2}\biggl(\frac{d{\bf A}^a}{dt}\cdot
\frac{d{\bf A}^a}{dt}\,+\,\omega_p^2\,{\bf A}^a\cdot {\bf A}^a\biggr)\,+\,
\frac{g^2}{4}\,f^{abc}\,f^{ade}\Bigl({\bf A}^b\cdot {\bf A}^d\Bigr)\,
\Bigl({\bf A}^c\cdot {\bf A}^e\Bigr).\eeq
 Eq.~(\ref{thetat}), together with the other
components of the energy-momentum tensor, can  also be obtained from
 $\Theta^{\mu\nu}_{{\rm sym}}$ given in Sec.3.3. Indeed,
for $A^\mu_a(x) = (0, {\bf A}^a(t))$, eqs.~(\ref{deps}) and (\ref{thetas})
 imply \beq\label{tsymt}
\Theta^{\mu\nu}_{\rm sym}&=&
3\,\omega_p^2\,\int\frac{d\Omega}{4\pi}\,v^\mu\,v^\nu\,\left\{
2\Bigl({\bf v}\cdot{\bf A}^a\Bigr)\Bigl({\bf v}\cdot{\bf A}^a\Bigr)\,-\,
\frac{1}{2}\,{\bf A}^a\cdot {\bf A}^a\right\},\eeq
and it is easily verified that  $\Theta_{\rm sym}^{00}=\Theta^{00}$.
Furthermore, $\Theta_{\rm sym}^{0i}=0$ and
\beq\label{tsymij}
\Theta^{ij}_{\rm sym}&=&
\omega_p^2\,\left\{
\frac{4}{5}\,A^i_a\,A^j_a\,-\,\frac{\delta^{ij}}{10}\,
\,{\bf A}^a\cdot {\bf A}^a\right\}.\eeq

 The external color sources which are needed to generate
the field configuration can be determined from the equations (\ref{ava})
which take here the form
\beq\label{rhot}
\rho (t)&=&i\,g\,\biggl[A^i,\frac{d A^i}{dt}\biggr],\eeq
and \beq\label{jet}
j^i(t)&=&\frac{d^2 A^i}{d t^2}\,+\,\omega^2_p A^i\,+\,g^2\,\biggl [\,
\Bigl[ A^i, A^j\Bigr],\,A^j\,\biggr].\eeq
Note that these equations differ
from the corresponding classical Yang-Mills equations in the vacuum only by the
presence of a mass term $\sim \omega^2_p \,A^i_a$
 for the vector potential.

In what follows, we  look
for self-sustained global color oscillations, i.e., solutions to the
equations (\ref{rhot})--(\ref{jet}) with vanishing external sources
($\rho(t)=j^i(t)=0$). The corresponding equations in the vacuum have been
extensively investigated, mostly for the $SU(2)$ color
group\cite{Baseyan79}. However, as we shall see,
the presence of the  mass term $\omega_p\sim gT$
dramatically affects the dynamical content of these equations.
Besides, the respective solutions for the high temperature quark-gluon plasma
have direct physical relevance, as they represent possible collective
excitations of the system. The physical interpretation of the
corresponding excitations in the confining regime is more obscure.

The condition $\rho(t)\,=\,0$ implies, according to eq.~(\ref{rhot}),
the commutativity of the color matrices $A^i(t)\equiv A^i_a(t) T^a$ and
${\dot A}^i(t)\equiv (d A^i_a/dt)T^a$. We shall consider here two non-trivial
possibilities:\\
({\bf i}) The  vectors $\{A^i_a(t)\}$
and $\{{\dot A}^i_a(t)\}$ are parallel in
color space for any $i$, which implies
$A^i_a(t)\,=\,{\cal A}^i_a\,h^i(t)$ (no summation over $i$),
 with constant ${\cal A}^i_a$ and arbitrary functions $h^i(t)$.\\
({\bf ii}) The color matrices $A^i(t)$ belong to the  Cartan subalgebra
of $SU(N)$, $A^i(t)\equiv \sum_s A^i_s(t) T^s$. (Recall the discussion at the
beginning of Sec.4.1.) If $N=2$,  ({\bf ii})
is just a particular case of ({\bf i}).

We consider now the consequences of the second set of equations,
that is,  $j^i(t)\,=\,0$  for all $i$, on
 the two possibilities mentioned  above.

({\bf i}) For gauge potentials of the  form
 $A^i(t)\,=\,{\cal A}^i\,h^i(t)$,
(${\cal A}^i\equiv{\cal A}^i_a T^a$),  the normal modes equations
  $j^i\,=\,0$ give generally non-linear and coupled
 second-order differential  equations. As an  example, consider
$SU(2)$, where $f^{abc} = \epsilon^{abc}$, ($a =1,\,2,\,3$),
 and the particular field configuration in which
${\cal A}_a^i=\delta^i_a$. Then, the functions $h_i(t)$ satisfy
\beq\label{h1}
\frac{d^2 h_1}{d t^2}\,+\,\omega^2_p\, h_1\,+\,g^2\,h_1\,\bigl(
h_2^2+h_3^2\bigr)\,=\,0,\eeq
plus two similar equations for $h_2(t)$ and  $h_3(t)$.
The associated  energy density,
\beq T^{00}&=&\frac{1}{2}\sum_i\left(\left(\frac{dh_i}{dt}\right )^2+
\omega_p^2\,h_i^2\right)\,+\,
\frac{g^2}{2}\,\Bigl(h_1^2\,h_2^2\,+\,h_1^2\,h_3^2\,
+h_2^2\,h_3^2\Bigr),\eeq
is an integral of the motion and acts as an effective Hamiltonian for the
functions $h_i(t)$. Contrary to what happens at zero temperature, where
$\omega_p=0$, here the energy conservation prevents any trajectory $\{h_i(t)\}$
from getting too far away from the origin in the 3-dimensional space with
axis $h_i$.

Periodic solutions to eqs.~(\ref{h1}) can be easily written down in the
particular case where all the three colors oscillate in phase:
$h_1=h_2=h_3\equiv h$ (``white oscillations''). The function $h(t)$
satisfies then the non-linear equation
\beq\label{h}
{\ddot h}\,+\,\omega^2_p\, h\,+2\,g^2\,h^3\,=\,0,\eeq
which has the following solution
\beq\label{hsol}
h(t)\,=\,h_\theta\,{\rm cn}\Bigl(\Omega_\theta (t-t_0); k\Bigr),\eeq
for the  initial conditions $h(t_0) = h_\theta$ and ${\dot h}(t_0) =0$
(the overdots indicate time derivatives).
Here,  ${\rm cn}(x;k)$ is the Jacobi elliptic cosine of argument $x$ and
modulus $k$, and $t_0$ is the arbitrary origin of the time. The
quantities $k$, $h_\theta$ and $\Omega_\theta$ are related
to the dimensionless parameter $\theta^2\equiv (g^2/\omega_p^4)T^{00}$  by
\beq\label{k}
k\,=\,\frac{1}{\sqrt 2}\,\biggl[1\,-\,\Bigl(1+\frac{8}{3}\,\theta^2\Bigr)
^{-1/2}\biggr]^{1/2},\eeq
\beq\label{htheta}
h_\theta\,=\,\frac{\omega_p}{{\sqrt 2}g}\,
\biggl[\Bigl(1+\frac{8}{3}\,\theta^2 \Bigr)^{1/2}\,-\,1\biggr]^{1/2},\eeq
and \beq\label{oth}
\Omega_\theta\,=\,\omega_p\,\Bigl(1+\frac{8}{3}\,\theta^2\Bigr)^{1/4}.\eeq
The solution (\ref{hsol}) is periodic, with period
\beq\label{T}
{\cal T}_\theta\,=\,\frac{4}{\Omega_\theta}\,K(k),\eeq
where $K(k)$ is the complete elliptic integral of modulus k. Since
$|h(t)|\simle T$, then $\theta\simle 1$ and  ${\cal T}_\theta$ is of order of
 ${\cal T}_0\equiv 2\pi/\omega_p$.
The associated field strenghts are
 $E^i_a=-\delta^i_a\,{\dot h}$,
and $B^i_a=g\,h^2\,\delta^i_a$. Therefore, the  vectors ${\bf E}_a$ and
  ${\bf B}_a$ are parallel for any $a$.

Other non-abelian solutions to eqs.~(\ref{h1}) are discussed
in Ref.~\cite{NAE}.

({\bf ii}) For gauge fields in the Cartan algebra,
 $A^i(t)\equiv \sum_s A^i_s(t) T^s$, the commutator entering eq.~(\ref{jet})
is trivially zero, so that the normal mode equation reduces to
\beq
\frac{d^2 A^i_s}{d t^2}\,+\,\omega^2_p A^i_s\,=\,0.\eeq
Its general solution has the form \beq\label{car}
A^i_s(t)\,=\,{\cal A}^i_s \cos\omega_p t\,+\, {\cal B}^i_s\sin\omega_p t,\eeq
with arbitrary constants ${\cal A}^i_s$ and ${\cal B}^i_s$.
A particular solution of this type for  $SU(3)$ is
${\bf A}(t)\,=\,T^3\,{\bf a}\cos\omega_p t\,+\,T^8\, {\bf b}\sin\omega_p t$,
where
${\bf a}$ and ${\bf b}$ are fixed vectors in coordinate space.
This describes coupled oscillations in both coordinate
and color spaces, with an energy density
\beq T^{00}\,=\,\frac{1}{2}\,\omega_p^2\,\Bigl({\bf a}^2\,+\,{\bf b}^2).\eeq
In particular,  $a^i = b^i$ corresponds to
global color rotations in the plane $T^3-T^8$.

\setcounter{equation}{0}
\section{Conclusions}

In this work we have studied the energy and the momentum of a general
soft gauge field configuration describing collective excitations of the
quark-gluon plasma. Our derivation relies on the explicit formulae
that we have obtained previously for the induced current expressing
the plasma response to the
 soft gauge fields. The simplicity of these expressions
allowed us to construct three different, but physically equivalent,
versions for the tensor $T^{\mu\nu}$,
which are gauge-invariant and rather simple.
 One of these versions is symmetric (and traceless),
which allowed us to simply obtain the gauge field angular momentum.
We have shown
that the total excitation energy is positive, for arbitrary gauge fields.
We have also verified that, for particular abelian field configurations,
our general expressions reduce to well-known formulae from classical
plasma physics. Finally, we have analyzed the global color dynamics of
the plasma and given a particular  non-linear solution which describes
a truly non-abelian excitation (the plasma white oscillations).
The specific examples that we have considered in the last two sections
 illustrate,
in particular, the utility of our general expressions for $T^{\mu\nu}$ for
various applications.

\setcounter{equation}{0}
\vspace*{2cm}
\renewcommand{\theequation}{A.\arabic{equation}}
\appendix{\noindent {\large{\bf Appendix}}}

In this appendix we discuss  the  leading order
 effective  propagator for soft gluons, together with  some related topics,
like the dielectric permitivity tensor for gauge plasmas and the polarization
vectors for the gluonic normal modes.

As already discussed in Sec. 2, the dominant contribution to the polarization
tensor for soft gauge fields is given by  $\Pi_{\mu\nu}(P)$,  eq.~(\ref{P}),
and satisfies $P^\mu\,\Pi_{\mu\nu}(P)=0$,
consequence of the conservation law (\ref{jcons}).
The soft gluon propagator is therefore ${}^*D_{\mu\nu}^{-1}(P)
\equiv D_{0\,\mu\nu}^{-1}(P)\,+\,\Pi_{\mu\nu}(P)$
 \cite{Pisarski,Braaten90b}.
In order to invert this expression, it is useful to decompose $\Pi_{\mu\nu}(P)$
into longitudinal and  transverse components (relative
to ${\bf p}$)\cite{Weldon82}:
\beq\label{PLT}
\Pi^{\mu\nu}(P)\equiv \Pi_l(p^0,p) {\cal P}_l^{\mu\nu}\,+\,
\Pi_t(p^0,p) {\cal P}_t^{\mu\nu},\eeq
where the projection operators ${\cal P}_{l,\,t}^{\mu\nu}$ satisfy
\beq\label{PLPT}
P_\mu\,{\cal P}_l^{\mu\nu}=P_\mu\, {\cal P}_t^{\mu\nu}=0,
\qquad\,\qquad {\cal P}_l^{\mu\nu}\,+
\,{\cal P}_t^{\mu\nu}\,=\,({P^\mu P^\nu}/{P^2})-g^{\mu\nu},\nonumber\\
{\cal P}_l^{\mu\rho}\, {\cal P}_{l\,\rho\nu}=-\,{\cal P}^{\mu}_{l\,\nu},\qquad
{\cal P}_t^{\mu\rho}\, {\cal P}_{t\,\rho\nu}=-\,{\cal P}^{\mu}_{t\,\nu},\qquad
{\cal P}_l^{\mu\rho}\, {\cal P}_{t\,\rho\nu}\,=\,0,\eeq
with $P^2=p_0^2-{\bf p}^2$ and $\hat p^i =p^i/p$.
 The explicit forms of the functions
$\Pi_{l,\,t}(p^0,p)$ follow easily from eqs.~(\ref{PLT})--(\ref{PLPT}) and
(\ref{P}); they can be inferred from eqs.~(\ref{el})--(\ref{et}) below.
Then, the  gluon propagator is
\beq\label{effp}
{}^*D^{\mu\nu}(P)\,=\,-\,{\cal P}_l^{\mu\nu}\,\,{}^*\Delta_l(P)\,-\,
{\cal P}_t^{\mu\nu}\,\,{}^*\Delta_t(P)\,+\,\frac
{1}{\lambda}\,\frac{P^\mu P^\nu}{P^4},\eeq
in a covariant gauge with gauge fixing parameter $\lambda$ ($\lambda=1$
in Feynman gauge). Here
\beq\label{effd}
{}^*\Delta_l^{-1}(P)\,\equiv\,P^2\,-\,\Pi_l(p^0,p),\qquad
{}^*\Delta_t^{-1}(P)\,\equiv\,P^2\,-\,\Pi_t(p^0,p),\eeq
are the effective inverse propagators for longitudinal and,
 respectively, transverse  gluons. They  vanish on the mass-shell for
soft quasi-gluons, thus  defining the corresponding dispersion relations,
which we denote by $\pm \omega_s(p)$, with $s=l$ or $t$. The
 mass-shell residues are defined by the  relation
\beq
{}^*\Delta_s (\omega, p)\equiv \frac {1}{\omega ^2-p^2-\Pi_s
 (\omega, p)}\approx\frac{z_s (p)}{\omega ^2 -\omega_s ^2(p)},\eeq
where the approximate equality holds in the vicinity of the  pole.
It follows that \beq\label{zlt}
z_s^{-1} (p)\,=\,\frac{1}{2\omega_s(p)}\,\frac{d}{d\omega}\,
{}^*\Delta_s^{-1}(\omega, p)\Big |_{\omega_s(p)}.\eeq
As discussed in Sec.4.3, after eq.~(\ref{ewave}), the residues $z_s(p)$ are
 positive functions.

Let us construct now covariant gauge
  polarization vectors appropriate for the longitudinal
($\epsilon^\mu({{\bf p}}\,;0)$), and, respectively, for the
transverse ($\epsilon^\mu({{\bf p}}\,;\lambda),\,\,\,\lambda=1,\, 2$)
 soft normal modes.
 They satisfy
\beq {}^*D_{\mu\nu}^{-1}\big(\omega_\lambda(p), {\bf p}\big)\,
\epsilon^\nu({\bf p};\lambda)\,=\,0,\eeq
where $\omega_\lambda(p)$ is equal to $\omega_l(p)$ for $\lambda =0$ and,
respectively, to $\omega_t(p)$ for $\lambda =1,\,2$.
By writting $\epsilon^\mu({\bf p};\lambda)=
(0, \bfepsilon({\bf p};\lambda))$, it follows that
$P_\mu\,\epsilon^\mu({{\bf p}};\lambda)=0$, for
$\lambda=0,1,2\,$ and $p^0=\omega_\lambda(p);\,$ furthermore,
${{\bf p}}\cdot{\bfepsilon}({{\bf p}};\lambda)=0$, for
$\lambda=1,2,\,$ and
 $(\delta^{ij}-\hat p^i\hat p^j)\epsilon^j({{\bf p}};0)=0.\,\,$
With the normalization $\epsilon ({{\bf p}};
\lambda)\cdot \epsilon^* ({{\bf p}};\lambda^\prime)=
-\,\delta_{\lambda\lambda^\prime}$, we obtain
\beq
\epsilon^\mu({{\bf p}};0)\,=\,\beta(p)\,(1,\,{\bf\hat p}\,\omega_l/p),\qquad
\beta(p)\beta^*(p)\,=\,p^2/(\omega_l^2-p^2),\eeq
and
\beq\label{pol}
\epsilon^\mu({{\bf p}};\lambda)\,=\,(0, \bfepsilon({\bf p};\lambda)),
\qquad {{\bf p}}\cdot{\bfepsilon}\,({{\bf p}};\lambda)=0,\qquad
\bfepsilon\,({{\bf p}};\lambda)\cdot \bfepsilon^{\,*}
\,({{\bf p}};\lambda^\prime)\,
=\,\delta_{\lambda\lambda^\prime},\eeq
for $\lambda=1,\,2$. Note also that
\beq
\sum_{\lambda=1,2}
\epsilon^\mu({{\bf p}};\lambda)\,\epsilon^{\nu\,*}({{\bf p}};\lambda)\,=\,
{\cal P}_t^{\mu\nu},\qquad \epsilon^\mu({{\bf p}};0)\,
\epsilon^{\nu\,*}({{\bf p}};0)\,=\,{\cal P}_l^{\mu\nu}.\eeq

The soft gluon effective propagator (\ref{effp}) is intimately related to the
{\it dielectric permitivity
tensor}, commonly used in classical plasma physics to
describe the polarization properties of the plasma.
In the weak field (or the abelian) limit, where the plasma responds linearly,
(i.e., $j^{ind}_\mu=\Pi_{\mu\nu}\,A^\nu$), we can use
 the standard definition of $\epsilon^{ij}$\cite{PhysKin}.
The {\it polarization
vector} ${{\bf P}}(\omega,\,{{\bf p}})$ satisfies, by definition,
$-i\omega\,{{\bf P}} (\omega,\,{{\bf p}})
\equiv {\bf j}^{\,ind}(\omega,\,{{\bf p}})$, and also
$i{{\bf p}}\cdot {{\bf P}}(\omega,\,{{\bf p}})=-\rho^{ind}
(\omega,\,{{\bf p}})$, because
of the continuity equation for $j_\mu^{ind}$, eq.~(\ref{jcons}). The
linear relation between  the {\it displacement vector}
${\bf D}(\omega,\,{{\bf p}})\equiv {\bf E}(\omega,\,{{\bf p}})\,+\,
{{\bf P}}(\omega,\,{{\bf p}})$  and
the electric field defines the dielectric tensor:
\beq D^i(\omega,\,{{\bf p}})\equiv \epsilon ^{ij}(\omega,\,{{\bf p}})\,
E^j(\omega,\,{{\bf p}}).\eeq
It follows immediately that
\beq\label{eps}\epsilon ^{ij}(\omega,\,{{\bf p}})\,=\,
\delta^{ij}\,-\,\frac{1}{\omega^2}\,\Pi^{ij}(\omega,\,{{\bf p}}).\eeq
For isotropic dielectric media, $\epsilon ^{ij}$ has only two independent
components, and it is useful to choose them as the   longitudinal and
the transverse dielectric functions, defined by
\beq\label{elt}
\epsilon ^{ij}(\omega,\,{{\bf p}})\,=\,\epsilon_l(\omega,p)\,
{\hat p^i\hat p^j}+
\epsilon_t(\omega,p)\,\left (\delta^{ij}\,-\,{\hat p^i \hat p^j}\right ).
\eeq
{}From  this relation, together with eqs.~(\ref{eps}) and (\ref{P}),
it follows that
\beq\label{el}
\epsilon_{l}(\omega, p)\,=\,1\,-\,\frac{1}{\omega^2-p^2}\,\Pi_l
(\omega, p)\,=\,1\,-\,3\,\frac{\omega_p^2}{p^2}\,\Bigl[
Q({\omega}/{p})\,-\,1\Bigr],\eeq
\beq\label{et}
\epsilon_{t}(\omega, p)\,=\,1\,-\,\frac{1}{\omega^2}\,\Pi_t
(\omega, p)
\,=\,1\,-\,\frac{3}{2}\,\frac{\omega_p^2}{p^2}\,\left [
1\,-\,\frac{\omega^2 - p^2}{\omega^2}\, Q({\omega}/{p})\right ],\eeq
where
\beq\label{Q}
Q(x)\equiv \frac{x}{2}\,\ln\frac{x+1}{x-1}\,=\,
\frac{x}{2}\,\left (\ln \Big |\frac{x+1}{x-1}\Big |\,-\,i\,\pi\theta
\,(1-|x|)\right ), \eeq
and the prescription $x\to x+i\eta$, $\eta \to 0_+$, has been understood in
writting the imaginary part of $Q(x)$.
 Remark the imaginary parts in $\epsilon_{l,\,t}
(\omega, p)$, which occur for $|\omega| < p$, as expected.
{}From eqs.~(\ref{el}), (\ref{et}) and (\ref{disp}), we derive
\beq\label{lres}
\frac {d}{d \omega}\left (\omega\,\epsilon_l (\omega,p)\right )\Big |_{\omega
_l(p)}\,=\,\omega_l(p)\,\frac{d \epsilon_l}{d \omega}\Big |_{\omega_l(p)}\,=\,
\frac{3\omega_p^2}{\omega_l^2(p)-p^2}\,-\,1,\eeq
and, similarly,
\beq\label{tres}
\frac{d}{d\omega}\left
[\omega\left (\epsilon_t(\omega,p)-p^2/\omega^2 \right )
\right ]_{\omega_t(p)}
&=&\omega_t(p)\,\frac{d}{d\omega}\left
(\epsilon_t(\omega,p)-p^2/\omega^2\right )\Big |_{\omega_t(p)}\nonumber\\
&=& \frac{3\omega_p^2}{\omega_t^2(p)-p^2}+\frac{p^2}{\omega_t^2
(p)}\,-\,1.\eeq
The dielectric functions are related to the inverse effective propagators,
as shown by eqs.~(\ref{effd}) and  (\ref{el})--(\ref{et}),
\beq \label{epsres}{}^*\Delta_l^{-1}(\omega,p)
=(\omega^2-p^2)\,\epsilon_l (\omega,p),\qquad\,\,
{}^*\Delta_t^{-1}(\omega,p)=\omega^2\,\epsilon_t (\omega,p)-p^2.\eeq
{}From the defining  relation of the residues, eq.~(\ref{zlt}),
it  follows that
\beq\label{eres}
  \omega_l(p)\,\frac{d \epsilon_l}{d \omega}\Big |_{\omega_l(p)}\,=\,
\frac{2\omega_l^2(p)}{\omega_l^2(p)-p^2}\,z_l^{-1}(p),\nonumber\\
\omega_t(p)\,\frac{d}{d\omega}\left
(\epsilon_t(\omega,p)-p^2/\omega^2\right )\Big |_{\omega_t(p)}
\,=\,2\,z_t^{-1}(p).\eeq

Note finally the  following angular integrals, which are needed
in Sec.4.3 in order to
 compute the energy and the momentum of the plasma waves:
\beq\label{vivj2}
\int\frac {d\Omega}{4\pi}\,\,\frac {v^i\,v^j}
{(\omega - {\bf v}\cdot{{\bf p}})^2}\,=\,\frac{ a({\omega}/{p})}{p^2}\,
\hat p^i\,\hat p^j\,+\,\frac{b({\omega}/{p})}{p^2}\,
(\delta^{ij}-\hat p^i\,\hat p^j),\eeq
and
\beq\label{vivj3}
\int\frac {d\Omega}{4\pi}\,\,\frac {v^i\,v^j}
{(\omega - {\bf v}\cdot{{\bf p}})^3}\,=\,\frac{ c({\omega}/{p})}{p^3}\,
\hat p^i\,\hat p^j\,+\,\frac{d({\omega}/{p})}{p^3}\,
(\delta^{ij}-\hat p^i\,\hat p^j).\eeq
These results are valid for $\omega > p$. In these formulae
\beq\label{ab}
{a}(x)\equiv 1\,+\,\frac{x^2}{x^2-1}\,-\,2Q(x),\qquad\,\,
{b}(x)\equiv Q(x)\,-\,1,\eeq
and
\beq\label{cd}
{c}(x)\equiv \frac{1}{x}\left \{Q(x)\,-\,\left (1-\frac{1}{x^2-1}\right )
\frac{x^2}{x^2-1}\right \},\qquad {d}(x)\equiv \frac{1}{2x}
\left (\frac{x^2}{x^2-1}\,-\,Q(x)\right ),\eeq
with $Q(x)$ given by (\ref{Q}).
For $x>1$, the functions $a(x)$, $b(x)$,  $c(x)$ and $d(x)$
are all positive and vanish  only in the limit $x\to \infty$.

\end{document}